\documentclass[eqsecnum,nofootinbib,11pt]{revtex4}
\usepackage{amsmath}
   \usepackage{bm}
\usepackage{amsthm,amscd}
\usepackage{mathrsfs}
\usepackage{verbatim}
\usepackage{amsfonts,amsmath,latexsym,amssymb,txfonts,bm}
\usepackage[american]{babel}
\usepackage{dsfont}
\usepackage[dvips]{graphicx}

\newcommand\bea{\begin{eqnarray}}
\newcommand\eea{\end{eqnarray}}
\newcommand{\diff}{\mathrm{d}}

\begin{document}
\title{The Casimir Effect for Generalized Piston Geometries}
\author{
Guglielmo Fucci\footnote{Electronic address: Guglielmo\textunderscore Fucci@Baylor.edu} and Klaus Kirsten\footnote{Electronic address: Klaus\textunderscore Kirsten@Baylor.edu}
\thanks{Electronic address: gfucci@nmt.edu}}
\affiliation{Department of Mathematics, Baylor University, Waco, TX 76798 USA
}
\date{\today}
\vspace{2cm}
\begin{abstract}

In this paper we study the Casimir energy and force for generalized pistons constructed from
warped product manifolds of the type $I\times_{f}N$ where $I=[a,b]$ is an interval of the real line and $N$ is a smooth compact Riemannian manifold
either with or without boundary. The piston geometry is obtained by dividing the warped product manifold into two regions separated
by the cross section positioned at $R\in(a,b)$. By exploiting zeta function regularization techniques we provide formulas for the Casimir energy and
force involving the arbitrary warping function $f$ and base manifold $N$.

\end{abstract}
\maketitle

\section{Introduction}	
The influence that external conditions like electro-magnetic and gravitational fields, non-trivial topology and the presence of boundaries have on the Casimir effect has been a very actively considered subject for several decades \cite{cas,bordag01,bordag09,elizalde,elizalde94,milton01,kim,kirsten01,dowk78,banach,gerald,david1,david2,larry}.
Whereas most research focuses on the impact one particular influence has, due mainly to technical progress the combined effect of several aspects is also being analyzed. In particular, flat or curved boundaries placed in a curved space-time have received considerable attention recently \cite{saha05-724-406,seta05-620-111,saha03-552-119,1109.1497}.
Examples are piston configurations, \cite{kirs09,teo09,eli09,iver,dowk11}
which have the advantage of rendering the Casimir force finite, \cite{cavalcanti04} although in a curved space-time this is not a generic feature anymore \cite{fucci11b,fucci11c}. Due to a different geometry in both chambers a complete cancelation of singularities in the force does not necessarily occur and ambiguities might remain. However, under certain conditions unambiguous answers can still be found as was shown recently for conical pistons \cite{fucci11b,fucci11c}.

The current article is a continuation of \cite{fucci11b,fucci11c}, where the conical manifold is replaced by a more general warped manifold. To analyze the Casimir energy and force we will use the zeta function regularization\cite{dowk76-13-3224,hawk77-55-133} and in Section 2 we describe the relevant spectral problem and the geometry of the piston. In Section 3 we find the analytical continuation of the associated zeta function; part of the construction involves the derivation of uniform asymptotics of solutions of initial value problems of an ordinary differential equation using the WKB approximation \cite{bend10b,mill06b}. The analytical continuation of the zeta function is the basis for the Casimir energy and force results in Section 4. Although our focus is the force on the piston, we present results for the energy for each chamber from which the force on the bounding plates of the chambers can be derived. Combining both chambers significant simplifications occur. In the Conclusions we summarize our results and outline possible further studies along the lines of our article.

\section{The Generalized Piston Geometry and Zeta Function}

Let $I=[a,b]\subset\mathbb{R}$ and $N$ be a $d$-dimensional smooth, compact Riemannian manifold either with or without boundary $\partial N$.
We consider the bounded manifold $M$ of dimension $D=d+1$ constructed as a warped product $M=I\times_{f}N$ assuming that
the warping function $f(r)>0$ for $r\in I$ and that $f\in C^{\infty}(I)$. The local geometry of the warped product manifold $M$ is described by the line element
\begin{equation}\label{0}
  \diff s^{2}=\diff r^{2}+f^{2}(r)\diff\Sigma^{2}\;,
\end{equation}
where $\diff\Sigma^{2}$ represents the line element on the manifold $N$.
It is possible to construct a piston configuration modeled
by the warped manifold $M$ as follows (cf. \cite{fucci11b,fucci11c}): Let $N_{R}$ be the cross section of the manifold $M$ positioned at the point $R\in(a,b)$.
The manifold $N_{R}$ divides $M$ into two separate regions denoted by region $I$ and region $II$. Region $I$ and region $II$
are represented, respectively, by the $D$-dimensional compact manifolds $M_{I}=[a,R]\times_{f}N$ and $M_{II}=(R,b]\times_{f}N$ having boundary
$\partial M_{I}=N_{a}\cup N_{R}$ and $\partial M_{II}=N_{R}\cup N_{b}$. It is clear from this construction that the warped product manifold $M$ is given
by the union of $M_{I}$ and $M_{II}$ along their common boundary $N_{R}$, namely $M=M_{I}\cup_{N_{R}}M_{II}$. The configuration described is a generalized piston where
the piston itself is modeled by the cross section $N_{R}$.

In this work we use zeta function regularization techniques \cite{dowk76-13-3224,hawk77-55-133} in
order to obtain the Casimir energy and force for the generalized piston configuration. For this reason we are interested in the analysis of the following eigenvalue
equation
\begin{equation}\label{2}
-\Delta_{M}\phi=\alpha^{2}\phi\;,
\end{equation}
where $\Delta_{M}$ represents the Laplace operator acting on scalar functions
$\phi\in\mathcal{L}^{2}(M)$.
On a manifold with metric (\ref{0}), $\Delta_{M}$ is expressed as
\begin{equation}\label{1}
  \Delta_{M}\phi=\left(\frac{\diff^{2}}{\diff r^{2}}+d\frac{f'(r)}{f(r)}\frac{\diff}{\diff r}+\frac{1}{f^{2}(r)}\Delta_{N}\right)\phi\;,
\end{equation}
where $\Delta_{N}$ is the Laplace operator on the manifold $N$.
Since the fields in one region are independent from the fields in the other region, the two spectral problems are
independent. This implies that we have a set of eigenvalues for each region, namely $\alpha_{I}$ for region $I$ and $\alpha_{II}$ for region $II$ with corresponding eigenfunctions $\phi_{I}$ and $\phi_{II}$. In both regions a solution to the eigenvalue equation (\ref{2}) can be written as a product $\phi_{p}=u_{\alpha_{p}}(r,\nu)\Phi(\omega)$, with $p$ denoting either $I$ or $II$, where
$\Phi(\omega)$ are the harmonics on $N$ satisfying
\begin{equation}\label{3}
-\Delta_N \Phi(\omega)=\nu^2
\Phi(\omega)\;,
\end{equation}
and $u_{\alpha_{p}}(r,\nu)$ denotes the solution of
\begin{equation}\label{4}
  \left(\frac{\diff^{2}}{\diff r^{2}}+d\frac{f'(r)}{f(r)}\frac{\diff}{\diff r}+\alpha_{p}^{2}-\frac{\nu^{2}}{f^{2}(r)}\right)u_{\alpha_{p}}(r,\nu)=0\;.
\end{equation}

The previous remarks suggest that the spectral zeta function associated with the piston configuration acquires contributions from the spectral problems in both regions and can, therefore, be written in the form
\begin{equation}\label{9a}
\zeta_{M}(s)=\zeta_{I}(s)+\zeta_{II}(s)\;,
\end{equation}
where $\zeta_{p}(s)$, with $p=(I,II)$, is the spectral zeta function corresponding to either region $I$ or $II$ defined as
\begin{equation}
  \zeta_{p}(s)=\sum\alpha_{p}^{-2s}\;.
\end{equation}
The explicit dependence of (\ref{4}) on the eigenvalues $\nu^2$ of $-\Delta_{N}$ suggests that the spectral problems on the warped manifold $M$
and on the manifold $N$ are closely related. For this reason $\zeta_{M}(s)$ will be expressed in terms of the spectral zeta function of the cross
section $N$ \cite{bordag96,cheeger83}
\begin{equation}
  \zeta_{N}(s)=\sum_{\nu} d(\nu)\nu^{-2s}\;,
\end{equation}
where $d(\nu)$ represents the finite degeneracy of each eigenvalue.
From the spectral zeta function (\ref{9a}) the Casimir energy is obtained as follows \cite{bordag01,bordag09,elizalde,elizalde94,kirsten01}
\begin{equation}\label{0a}
E_{\textrm{Cas}}=\lim_{\varepsilon\to 0}\frac{\mu^{2\varepsilon}}{2}\zeta_{M}\left(\varepsilon-\frac{1}{2}\right)\;,
\end{equation}
where $\mu$ represents an arbitrary mass parameter. Since $\zeta_{M}(s)$ will generally present a pole at $s=-1/2$, performing the limit in (\ref{0a})
leads to the expression
\begin{equation}\label{41a}
E_{\textrm{Cas}}=\frac{1}{2}\textrm{FP}\zeta_{M}\left(-\frac{1}{2}\right)+\frac{1}{2}\left(\frac{1}{\varepsilon}+\ln\mu^{2}\right)\textrm{Res}\,\zeta_{M}\left(-\frac{1}{2}\right)+O(\varepsilon)\;,
\end{equation}
where $\textrm{Res}$ and $\textrm{FP}$ denote the residue and the finite part.
For piston configurations the Casimir energy depends explicitly on the position $R$ of the piston \cite{cavalcanti04} and the Casimir force
can be obtained through the relation
\begin{equation}\label{10a}
F_{\textrm{Cas}}(R)=-\frac{\partial}{\partial R}E_{\textrm{Cas}}(R)\;.
\end{equation}
It is not difficult to realize, from the expressions (\ref{41a}) and (\ref{10a}), that the Casimir force on the piston
is well defined {\it only} when $\textrm{Res}\,\zeta_{M}(-1/2)$ does not depend on the position $R$.

\section{Analytic Continuation of the Spectral Zeta Function}

The spectral zeta function in (\ref{9a}) is well defined in the region $\Re(s)>D/2$ and in order to compute the Casimir energy we
need to find its analytic continuation to a neighborhood of $s=-1/2$. Although the eigenvalues $\alpha_{i}$ are not explicitly known for a general warping function $f$,
they can be found implicitly once appropriate boundary conditions are imposed. For definiteness we impose Dirichlet boundary conditions in both regions $I$ and $II$, namely
\begin{equation}\label{5}
  u_{\alpha_{I}}(a,\nu)=u_{\alpha_{I}}(R,\nu)=0\;, \quad\textrm{and}\quad u_{\alpha_{II}}(R,\nu)=u_{\alpha_{II}}(b,\nu)=0\;.
\end{equation}
Instead of considering the boundary value problem
(\ref{4}) and (\ref{5}), it is convenient to analyze the following initial value problem \cite{kirs04-37-4649}
\begin{eqnarray}\label{4a}
  \left(\frac{\diff^{2}}{\diff r^{2}}+d\frac{f'(r)}{f(r)}\frac{\diff}{\diff r}+\rho_{p}^{2}-\frac{\nu^{2}}{f^{2}(r)}\right)u_{\rho_{p}}(r,\nu)=0\;,
\end{eqnarray}
with $\rho_{p} \in\mathbb{C}$,
\begin{eqnarray}\label{400}
  u_{\rho_{I}}(a,\nu)=0,\; u'_{\rho_{I}}(a,\nu)=1\;,
\end{eqnarray}
in region $I$, and
\begin{equation}\label{401}
  u_{\rho_{II}}(R,\nu)=0,\; u'_{\rho_{II}}(R,\nu)=1\;,
\end{equation}
in region $II$.
The two sets of eigenvalues $\alpha_{I}$ and $\alpha_{II}$ are then obtained implicitly
as solutions of the equations
\begin{equation}\label{6}
  u_{\rho_{I}}(R,\nu)=0\;, \qquad  u_{\rho_{II}}(b,\nu)=0\;.
\end{equation}

To avoid displaying similar explicit expressions valid in region $I$ and $II$ we will employ from now on the notation $x_{p}$ with the meaning $x_{I}=R$ and $x_{II}=b$. Bearing in mind the last remark and thanks to the relations (\ref{5}) the spectral zeta functions for region $I$ and $II$ can be rewritten in terms of the integral representation \cite{kirsten01}
\begin{equation}\label{7}
  \zeta_{p}(s)=\frac{1}{2\pi i}\sum_{\nu}\int_{\gamma_{p}}dk (k^{2}+m^{2})^{-s}\frac{\partial}{\partial k}\ln u_{k}(x_{p},\nu)\;,
\end{equation}
where $\gamma_{p}$ is a contour that encircles, in the counterclockwise direction, the real zeroes of $u_{k}(x_{p},\nu)$.
The spectral parameter $m$ has been introduced for technical reasons and it will be sent to zero in the final results.
By first deforming the contour $\gamma_{p}$ to the imaginary axis and by then performing the change of variable $k=z\nu$, the expression (\ref{7})
can be rewritten as
\begin{equation}\label{9}
  \zeta_{p}(s)=\sum_{\nu}d(\nu)\zeta_{p}^{\nu}(s)\;,
\end{equation}
where
\begin{equation}\label{10}
  \zeta^{\nu}_{p}(s)=\frac{\sin \pi s}{\pi}\int_{\frac{m}{\nu}}^{\infty}dz (\nu^{2}z^{2}-m^{2})^{-s}\frac{\partial}{\partial z}\ln u_{i\nu z}(x_{p},\nu)\;.
\end{equation}
The integral in (\ref{10}) is well defined for $1/2<\Re(s)<1$. In order to perform the analytic continuation
to a neighborhood of $s=-1/2$ we will add and subtract, from (\ref{10}), a suitable number of terms from the uniform asymptotic expansion, as $\nu\to\infty$ and $z=k/\nu$ fixed, of the functions $u_{i\nu z}(x_{p},\nu)$.

The desired uniform asymptotic expansion can be obtained by employing a WKB approximation \cite{fucci11}. The starting point of this method is the use of the
following expression for the solution of (\ref{4a})
\begin{equation}\label{11}
  u_{i\nu z}(r,\nu)=\exp\left\{-\frac{d}{2}\int_a^{r}U(t)\,\diff t\right\}\Psi_{\nu}(z,r)\;,
\end{equation}
with $U(t)=f'(t)/f(t)$.  Once (\ref{11}) is substituted into (\ref{4a}) we obtain
\begin{equation}\label{8}
  \left(\frac{\diff^{2}}{\diff r^{2}}+q(\nu,z,r)\right)\Psi_{\nu}(z,r)=0\;,
\end{equation}
with the introduction of the function
\begin{eqnarray}
  q(\nu,z,r)&=&V(\nu,z,r)-\frac{1}{2}U'(r)-\frac{1}{4}U^{2}(r)\nonumber\\
  &=&-\nu^{2}\left(z^{2}+\frac{1}{f^{2}(r)}\right)-\frac{d}{2}\frac{f''(r)}{f(r)}-\frac{d(d-2)}{4}\frac{{f'}^{2}(r)}{f^{2}(r)}\;.
\end{eqnarray}
In order to proceed with the WKB method, we define the auxiliary function
\begin{equation}\label{8a}
 \mathcal{S}(\nu,z,r)=\frac{\partial}{\partial r}\ln \Psi_{\nu}(z,r)\;,
\end{equation}
which can be easily shown to satisfy the non-linear differential equation
\begin{equation}\label{12}
  \mathcal{S}'(\nu,z,r)=-q(\nu,z,r)-\mathcal{S}^{2}(\nu,z,r)\;.
\end{equation}
The uniform asymptotic expansion for the function $\mathcal{S}(\nu,z,r)$ as $\nu\to\infty$, and hence for $\Psi_{\nu}(z,r)$ through (\ref{8a}), is obtained by substituting the ansatz
\begin{equation}\label{13}
  \mathcal{S}(\nu,z,r)\sim \nu\,S_{-1}(z,r)+S_{0}(z,r)+\sum_{i=1}^{\infty}\frac{S_{i}(z,r)}{\nu^{i}}\;,
\end{equation}
into (\ref{12}) and by equating like powers of $\nu$. This procedure leads to the following expressions for the functions $S_{i}(z,r)$ with $i=\{-1, 0, 1,\cdots\}$,
\begin{equation}\label{11a}
  S_{-1}^{\pm}(z,r)=\pm\sqrt{z^{2}+\frac{1}{f^{2}(r)}}\;,\qquad S_{0}(z,r)=-\frac{1}{2}\frac{\partial}{\partial r}\ln S^{\pm}_{-1}(z,r)\;,
\end{equation}
\begin{equation}\label{11aa}
  S^{\pm}_{1}(z,r)=-\frac{1}{2S^{\pm}_{-1}(z,r)}\left[-\frac{d}{2}\frac{f''(r)}{f(r)}-\frac{d(d-2)}{4}\frac{{f'}^{2}(r)}{f^{2}(r)}+S_{0}^{2}(z,r)+S'_{0}(z,r)\right]\;,
\end{equation}
and, for the higher asymptotic orders with $i\geq 1$,
\begin{equation}\label{12a}
  S_{i+1}^{\pm}(z,r)=-\frac{1}{2S^{\pm}_{-1}(z,r)}\left[{S'_{i}}^{\pm}(z,r)+\sum_{n=0}^{i}S^{\pm}_{n}(z,r)S^{\pm}_{i-n}(z,r)\right]\;.
\end{equation}
Depending on the choice of sign for the functions $S_{i}(z,r)$ one obtains either the exponentially decaying or the exponentially increasing part of the asymptotic expansion for $\Psi_{\nu}(z,r)$. The correct asymptotic expression for $u_{i\nu z}(r,\nu)$ contains a linear combination of both behaviors \cite{fucci11} and it reads
\begin{eqnarray}\label{14}
  u_{i\nu z}(r,\nu)&=&\exp\left\{-\frac{d}{2}\int_{a}^{r}U(t)\,\diff t\right\}\Bigg[A\exp\left\{\int_{a}^{r}\mathcal{S}^{+}(\nu,z,t)\diff t\right\}\nonumber\\
  &+&B\exp\left\{\int_{a}^{r}\mathcal{S}^{-}(\nu,z,t)\diff t\right\}\Bigg]\;,
\end{eqnarray}
where $A$ and $B$ are uniquely determined once the initial conditions are imposed.

By imposing the initial condition (\ref{400}) in region $I$,
(\ref{401}) in region $II$, and by disregarding exponentially small terms we obtain the expansions
\begin{eqnarray}\label{15}
   \ln u_{i\nu z}(R,\nu)&=&-\ln\left[\mathcal{S}^{+}(\nu,z,a)-\mathcal{S}^{-}(\nu,z,a)\right]-\frac{d}{2}\int_{a}^{R}U(t)\,\diff t\nonumber\\
   &+&\int_{a}^{R}\mathcal{S}^{+}(\nu,z,t)\diff t\;,
\end{eqnarray}
in region $I$ and
\begin{eqnarray}\label{16}
   \ln u_{i\nu z}(b,\nu)&=&-\ln\left[\mathcal{S}^{+}(\nu,z,R)-\mathcal{S}^{-}(\nu,z,R)\right]-\frac{d}{2}\int_{R}^{b}U(t)\,\diff t\nonumber\\
  & +&\int_{R}^{b}\mathcal{S}^{+}(\nu,z,t)\diff t\;,
\end{eqnarray}
in region $II$.

Let us observe at this point that when adding up these two expressions, most terms will become $R$-independent. Therefore no contribution to the force (\ref{10a}) on the piston will result.

For the first term on the left hand side of (\ref{15}) one has an expansion of the form
\begin{equation}\label{18}
  \ln\left[\mathcal{S}^{+}(\nu,z,a)-\mathcal{S}^{-}(\nu,z,a)\right]=\ln\left(2\nu\right)+\frac{1}{2}\ln\left(z^{2}+\frac{1}{f^{2}(a)}\,\right)+\sum_{i=1}^{\infty}\frac{\mathcal{D}_{i}(z,a)}{\nu^{i+1}}\;,
\end{equation}
where the $\mathcal{D}_{i}(z,a)$ are determined through the relation
\begin{equation}\label{18a}
  \ln\left[1+\frac{1}{2}\left(z^{2}+\frac{1}{f^{2}(a)}\right)^{-\frac{1}{2}}\sum_{k=1}^{\infty}\frac{\omega_{k}(z,a)}{\nu^{k+1}}\right]\simeq \sum_{i=1}^{\infty}\frac{\mathcal{D}_{i}(z,a)}{\nu^{i+1}}
\end{equation}
with $\omega_{i}(z,a)=S_{i}^{+}(z,a)-S_{i}^{-}(z,a)$. Obviously, the same expansion holds for the term $\ln\left[\mathcal{S}^{+}(\nu,z,R)-\mathcal{S}^{-}(\nu,z,R)\right]$ in (\ref{16}) once
$a$ is replaced with $R$.

By utilizing the asymptotic expansion (\ref{13}), the explicit expression for $U(t)$, and the result (\ref{18}) the expansion (\ref{15})
can be rewritten as
\begin{eqnarray}\label{20}
  \ln u_{i\nu z}(R,\nu)&=&-\ln\left(2\nu\right)-\frac{1}{2}\ln\left(z^{2}+\frac{1}{f^{2}(a)}\,\right)+\frac{1}{4}\ln\left[\frac{1+z^{2}f^{2}(a)}{1+z^{2}f^{2}(R)}\right]\nonumber\\
  &+&\frac{d-1}{2}\ln\frac{f(a)}{f(R)}+\nu\int_{a}^{R}S_{-1}^{+}(z,t)\diff t+\sum_{i=1}^{\infty}\frac{\mathcal{M}_{i}(z,a,R)}{\nu^{i}}\;,
\end{eqnarray}
where
\begin{equation}\label{20b}
  \mathcal{M}_{i}(z,a,R)=\int_{a}^{R}S_{i}^{+}(z,t)\diff t-\mathcal{D}_{i-1}(z,a)\;,
\end{equation}
for $i\geq 1$ and $\mathcal{D}_{0}(z,a)=0$. The asymptotic expansion for $\ln u_{i\nu z}(b,\nu)$ in region $II$ is obtained from (\ref{20})
once the replacement $a\to R$ and $R\to b$ is performed.

Since the process of analytic continuation of the spectral zeta function involves the computation of integrals with respect to the
variable $z$ of the terms of the uniform asymptotic expansion of the functions $u_{i\nu z} (x_p, \nu )$, it is convenient to render manifest the dependence on $z$
of the functions $S_{i}^{+}(z,t)$ and $\mathcal{M}_{i}(z,a,R)$. From the expression (\ref{13}) and the results (\ref{11a})-(\ref{12a}) one can prove that
the functions $S_{i}^{+}(z,t)$ have the general form
\begin{equation}\label{21}
  S^{+}_{i}(z,r)=\sum_{k=0}^{i+1}F_{k,\,i}(r)f^{2k+i}(r)\left(1+z^{2}f^{2}(r)\right)^{-\frac{2k+i}{2}}\;.
\end{equation}
From the expression for $S_{0}(z,r)$ and $S^{+}_{1}(z,r)$ in (\ref{11a}) and (\ref{11aa}) it is not difficult to obtain
\begin{equation}
  F_{0,\,0}(r)=0\;,\quad F_{1,\,0}(r)=\frac{f'(r)}{2f^{3}(r)}\;,\quad F_{0,\,1}(r)=\frac{d}{4}\frac{f''(r)}{f(r)}+\frac{d(d-2)}{8}\frac{{f'}^{2}(r)}{f^2(r)}\;,
\end{equation}
\begin{equation}
  F_{1,\,1}(r)=\frac{3}{4}\frac{ {f'}^{2}(r)}{ f^{4}(r)}-\frac{1}{4}\frac{f''(r)}{ f^{3}(r)}\;,\qquad F_{2,\,1}(r)=-\frac{5}{8}\frac{ {f'}^{2}(r)}{ f^{6}(r)}\;.
\end{equation}
The functions $F_{k,\,i}(r)$ of higher order are found by using the general form (\ref{21}) of $S_i^{+}(z,t)$
into the recurrence relation (\ref{13}). More explicitly one has, for $i\geq 1$,
\begin{eqnarray}\label{22}
  F_{k,\,i+1}(r)=-\frac{1}{2}\left[(2k+i-2)\frac{F_{k-1,\,i}(r)f'(r)}{f^{3}(r)}+F'_{k,\,i}(r)H(i+1-k)+\mathcal{K}_{k,\,i}(r)\right],\;\;\;\;\;\;
\end{eqnarray}
where $H(x)$ denotes the Heaviside function and $\mathcal{K}_{k,\,i}(r)$ is defined through
\begin{eqnarray}\label{23}
  \sum_{m=0}^{i}\,\,\sum_{k=0}^{m+1}\,\,\sum_{j=0}^{i-m+1}F_{k,\,m}(r)F_{j,\,i-m}(r)f^{2k+2j+i}(r)\left(1+z^{2}f^{2}(r)\right)^{-\frac{2k+2j+i}{2}}\nonumber\\
  =\sum_{n=0}^{i+2}\mathcal{K}_{n,\,i}(r)f^{2n+i}(r)\left(1+z^{2}f^{2}(r)\right)^{-\frac{2n+i}{2}}\;.
\end{eqnarray}

The functions $\mathcal{M}_{i}(z,a,R)$ defined in (\ref{20b}) contain the terms $\mathcal{D}_{i}(z,a)$ which, according to the cumulant expansion (\ref{18a}), have the form
\begin{equation}\label{24}
  \mathcal{D}_{2i-1}(z,a)=\sum_{k=0}^{2i}\Omega_{k,\,i}(a)f^{2k+2i}(a)\left(1+z^{2}f^{2}(a)\right)^{-k-i}\;,
\end{equation}
with $\Omega_{k,\,i}(a)$ found from the relation
\begin{eqnarray}\label{25}
  \ln\left[1+\sum_{j=1}^{\infty}\frac{1}{\nu^{2j}}\left(\sum_{n=0}^{2j}F_{n,\,2j-1}(a)f^{2n+2j}(a)\left(1+z^{2}f^{2}(a)\right)^{-n-j}\right)\right]\nonumber\\
  \simeq \sum_{i=1}^{\infty}\frac{1}{\nu^{2i}}\sum_{k=0}^{2i}\Omega_{k,\,i}(a)f^{2k+2i}(a)\left(1+z^{2}f^{2}(a)\right)^{-k-i}\;.
\end{eqnarray}
The relation (\ref{24}) together with (\ref{11aa}) and (\ref{12a}) allow us to derive the following expressions for $\mathcal{M}_{i}(z,a,R)$:
when $i=2m+1$ and $m\in\mathbb{N}_{0}$, we have
\begin{equation}\label{26}
  \mathcal{M}_{2m+1}(z,a,R)=\sum_{k=0}^{2m+2}\int_{a}^{R}\diff t \,F_{k,\,2m+1}(t)f^{2k+2m+1}(t)\left(1+z^{2}f^{2}(t)\right)^{-\frac{2k+2m+1}{2}}\;,
\end{equation}
while for $i=2m$ and $m\in\mathbb{N}^{+}$ we obtain
\begin{eqnarray}\label{27}
  \mathcal{M}_{2m}(z,a,R)&=&\sum_{k=0}^{2m+1}\int_{a}^{R}\diff t \,F_{k,\,2m}(t)f^{2k+2m}(t)\left(1+z^{2}f^{2}(t)\right)^{-k-m}\nonumber\\
  &-&\sum_{k=0}^{2m}\Omega_{k,\,m}(a)f^{2k+2m}(a)\left(1+z^{2}f^{2}(a)\right)^{-k-m}\;.
\end{eqnarray}
We would like to remind the reader that similar results for $\mathcal{M}_{i}(z,R,b)$ are obtained in region $II$ once the replacement $a\to R$ and $R\to b$ is made.

Since the relevant uniform asymptotic expansion is now completely known we can proceed with the analytic continuation of the spectral zeta function (\ref{9a}).
By adding and subtracting in (\ref{10}) $N=D$ terms of the uniform asymptotic expansion (\ref{20}) we obtain, in region $I$,
\begin{equation}\label{28}
  \zeta_{I}(s,a,R)=Z_{I}(s,a,R)+\sum_{i=-1}^{D}A^{(I)}_{i}(s,a,R)\;.
\end{equation}
The function $Z_{I}(s,a,R)$, in the massless case, has the following integral representation valid for $\Re(s)>-1$
\begin{eqnarray}\label{29}
  Z_{I}(s,a,R)&=&\frac{\sin\pi s}{\pi}\sum_{\nu}d(\nu)\nu^{-2s}\int_{0}^{\infty}\diff z z^{-2s}\frac{\partial}{\partial z}\Bigg\{\ln u_{i\nu z}(R,\nu)+\ln\left(2\nu\right)\nonumber\\
  &+&\frac{1}{2}\ln\left(z^{2}+\frac{1}{f^{2}(a)}\,\right)
  -\frac{1}{4}\ln\left[\frac{1+z^{2}f^{2}(a)}{1+z^{2}f^{2}(R)}\right]-\frac{d-1}{2}\ln\frac{f(a)}{f(R)}\nonumber\\
  &-&\nu\int_{a}^{R}S_{-1}^{+}(z,t)\diff t-\sum_{i=1}^{D}\frac{\mathcal{M}_{i}(z,a,R)}{\nu^{i}}\Bigg\}\;.
\end{eqnarray}
The contributions $A^{(I)}_{i}(s,a,R)$ are obtained, instead, by integrating with respect to the variable $z$ the asymptotic terms in (\ref{20})
and they read, in the massless case,
\begin{equation}\label{30}
  A_{-1}^{(I)}(s,a,R)=\frac{1}{2\sqrt{\pi}}\frac{\Gamma\left(s-\frac{1}{2}\right)}{\Gamma(s)}\zeta_{N}\left(s-\frac{1}{2}\right)\int_{a}^{R}f^{2s-1}(t)\diff t\;,
\end{equation}
\begin{equation}\label{31}
  A_{0}^{(I)}(s,a,R)=-\frac{1}{4}\zeta_{N}(s)\left[f^{2s}(a)+f^{2s}(R)\right]\;,
\end{equation}
\begin{eqnarray}\label{32}
  A^{(I)}_{2n+1}(s,a,R)&=&-\frac{1}{\Gamma(s)}\zeta_{N}\left(s+n+\frac{1}{2}\right)\sum_{j=0}^{2n+2}\frac{\Gamma\left(s+j+n+\frac{1}{2}\right)}{\Gamma\left(j+n+\frac{1}{2}\right)}\nonumber\\
  &\times&\int_{a}^{R}F_{j,\,2n+1}(t)f^{2(s+j+n)+1}(t)\diff t\;,
\end{eqnarray}
for $n\in\mathbb{N}_{0}$, and
\begin{eqnarray}\label{33}
  A^{(I)}_{2n}(s,a,R)&=&-\frac{1}{\Gamma(s)}\zeta_{N}\left(s+n\right)\sum_{j=0}^{2n+1}\frac{\Gamma\left(s+j+n\right)}{\Gamma\left(j+n\right)}\nonumber\\
  &\times&\left[\int_{a}^{R}F_{j,\,2n}(t)f^{2(s+j+n)}(t)\diff t-\Omega_{j,\,n}(a)f^{2(s+j+n)}(a)\right]\;,
\end{eqnarray}
when $n\in\mathbb{N}^{+}$ and where $\Omega_{2n+1,n}=0$. In region $II$ we obtain similar results for the analytic continuation of the spectral zeta function. In fact, one has
\begin{equation}\label{34}
  \zeta_{II}(s,R,b)=Z_{II}(s,R,b)+\sum_{i=-1}^{D}A^{(II)}_{i}(s,R,b)\;.
\end{equation}
The terms $Z_{II}(s,R,b)$ and $A^{(II)}_{i}(s,R,b)$ can be obtained from, respectively, (\ref{29}) and (\ref{30})-(\ref{33}) once we replace $a\to R$ and $R\to b$.

\section{The Casimir Energy and Force for Generalized Pistons}

In order to compute the Casimir energy in the framework of zeta function regularization we employ the relation (\ref{41a}).
By setting $s=\varepsilon-1/2$ and by taking into account the meromorphic structure of $\zeta_{N}(s)$ \cite{kirsten01} we obtain for the asymptotic terms (\ref{30})-(\ref{33}) in region $I$ the expansions
\begin{eqnarray}\label{35a}
  A_{-1}^{(I)}(\varepsilon-1/2,a,R)&=&\frac{1}{4\pi\varepsilon}\zeta_{N}(-1)\int_{a}^{R}f^{-2}(t)\diff t+\frac{1}{2\pi}\zeta_{N}(-1)\int_{a}^{R}f^{-2}(t)\ln f(t)\diff t\nonumber\\
  &+&\frac{1}{4\pi}\left[\zeta'_{N}(-1)+(\ln 4-1)\zeta_{N}(-1)\right]\int_{a}^{R}f^{-2}(t)\diff t+O(\varepsilon)\;,
\end{eqnarray}
\begin{eqnarray}\label{35}
  \lefteqn{A_{0}^{(I)}(\varepsilon-1/2,a,R)=-\frac{1}{4\varepsilon}\textrm{Res}\,\zeta_{N}\left(-\frac{1}{2}\right)\left[f^{-1}(a)+f^{-1}(R)\right]}\nonumber\\
  &&-\frac{1}{4}\textrm{FP}\,\zeta_{N}\left(-\frac{1}{2}\right)\left[f^{-1}(a)+f^{-1}(R)\right]\nonumber\\
  &&-\frac{1}{2}\textrm{Res}\,\zeta_{N}\left(-\frac{1}{2}\right)\left[f^{-1}(a)\ln f(a)+f^{-1}(R)\ln f(R)\right]+O(\varepsilon)\;,
\end{eqnarray}
and
\begin{eqnarray}\label{36}
  A_{1}^{(I)}(\varepsilon-1/2,a,R)&=&\frac{1}{2\pi\varepsilon}\zeta_{N}(0)\int_{a}^{R}F_{0,1}(t)\diff t+\frac{1}{2\pi}\zeta'_{N}(0)\int_{a}^{R}F_{0,1}(t)\diff t\nonumber\\
  &+&\frac{1}{2\pi}\zeta_{N}(0)\left[2\int_{a}^{R}F_{0,1}(t)\ln f(t)\diff t+(\ln 2-1)\int_{a}^{R}F_{0,1}(t)\diff t\right]\nonumber\\
  &-&\frac{1}{2\sqrt{\pi}}\zeta_{N}(0)\sum_{j=1}^{2}\frac{\Gamma(j)}{\Gamma\left(j+\frac{1}{2}\right)}\int_{a}^{R}F_{j,1}(t)f^{2j}(t)\diff t+O(\varepsilon)\;.
\end{eqnarray}
The remaining asymptotic terms with $n\geq 1$ have the form
\begin{eqnarray}\label{37}
  \lefteqn{ A_{2n+1}^{(I)}(\varepsilon-1/2,a,R)=\frac{1}{\sqrt{\pi}}\textrm{Res}\,\zeta_{N}\left(n\right)\sum_{j=0}^{2n+2}\frac{\Gamma(j+n)}{\Gamma\left(j+n+\frac{1}{2}\right)}
  \Bigg[\frac{\mathcal{A}_{j,n}(a,R)}{\varepsilon}}\nonumber\\
  &+&\mathcal{B}_{j,n}(a,R)\Bigg]
  +\frac{1}{\sqrt{\pi}}\textrm{FP}\,\zeta_{N}\left(n\right)\sum_{j=0}^{2n+2}\frac{\Gamma(j+n)}{\Gamma\left(j+n+\frac{1}{2}\right)}\mathcal{A}_{j,n}(a,R)+O(\varepsilon)\;,
\end{eqnarray}
and
\begin{eqnarray}\label{38}
  \lefteqn{A_{2n}^{(I)}(\varepsilon-1/2,a,R)=\frac{1}{\sqrt{\pi}}\textrm{Res}\,\zeta_{N}\left(n-\frac{1}{2}\right)\sum_{j=0}^{2n+1}\frac{\Gamma\left(j+n-\frac{1}{2}\right)}{\Gamma\left(j+n\right)}
  \Bigg[\frac{\mathcal{C}_{j,n}(a,R)}{\varepsilon}}\nonumber\\
  &&+\mathcal{F}_{j,n}(a,R)\Bigg]
  +\frac{1}{\sqrt{\pi}}\textrm{FP}\,\zeta_{N}\left(n-\frac{1}{2}\right)\sum_{j=0}^{2n+1}\frac{\Gamma\left(j+n-\frac{1}{2}\right)}{\Gamma\left(j+n\right)}\mathcal{C}_{j,n}(a,R)+O(\varepsilon)\;,
\end{eqnarray}
where for typographical convenience we have introduced the functions
\begin{equation}\label{39}
  \mathcal{A}_{j,n}(a,R)=\frac{1}{2}\int_{a}^{R}F_{j,\,2n+1}(t)f^{2j+2n}(t)\diff t\;,
\end{equation}
\begin{equation}\label{40}
  \mathcal{B}_{j,n}(a,R)=\int_{a}^{R}F_{j,\,2n+1}(t)f^{2j+2n}(t)\ln f(t)\diff t+\left(H_{n+j-1}-2+2\ln 2\right)\mathcal{A}_{j,n}(a,R)\,,
\end{equation}
\begin{equation}\label{41}
  \mathcal{C}_{j,n}(a,R)=\frac{1}{2}\int_{a}^{R}F_{j,\,2n}(t)f^{2j+2n-1}(t)\diff t-\mathcal{T}_{j,n}(a)\;,
\end{equation}
with
\begin{equation}\label{48aaa}
  \mathcal{T}_{j,n}(a)=\frac{1}{2}\Omega_{j,\,n}(a)f^{2j+2n-1}(a)\;,
\end{equation}
and
\begin{eqnarray}\label{42}
  \mathcal{F}_{j,n}(a,R)&=&\int_{a}^{R}F_{j,\,2n}(t)f^{2j+2n-1}(t)\ln f(t)\diff t-\Omega_{j,\,n}(a)f^{2j+2n-1}(a)\ln f(a)\nonumber\\
  &+&\sum_{k=0}^{n+j-1}\frac{2}{2k-1}\,\mathcal{C}_{j,n}(a,R)\;,
\end{eqnarray}
where $H_{n}$ represent the harmonic numbers.
In addition, since $Z_{I}(s,a,R)$ is an analytic function in the halfplane $\Re(s)>-1$ the value $s=-1/2$ can simply be set in (\ref{29}). Once again, the same results are obtained in region $II$ with the replacement indicated before.

From the results (\ref{35a})-(\ref{38}) it is not very difficult to compute the Casimir energy associated with region $I$.
In fact, by utilizing the formula (\ref{41a}), with the spectral zeta function $\zeta_{M}$ replaced by $\zeta_{I}$, we obtain the following expression
\begin{eqnarray}\label{new}
  E^{(I)}_{\textrm{Cas}}(a,R)&=&\frac{1}{2}Z_{I}\left(-\frac{1}{2},a,R\right)+\frac{1}{2}\mathcal{Q}(a,R)\nonumber\\
  &+&\frac{1}{2\sqrt{\pi}}\sum_{n=1}^{\left[\frac{D-1}{2}\right]}\sum_{j=0}^{2n+2}\frac{\Gamma(j+n)}{\Gamma\left(j+n+\frac{1}{2}\right)}
  \big[\textrm{FP}\,\zeta_{N}\left(n\right)\mathcal{A}_{j,n}(a,R)\nonumber\\
  &+&\textrm{Res}\,\zeta_{N}\left(n\right)\mathcal{B}_{j,n}(a,R)\big]\nonumber\\
  &+&\frac{1}{2\sqrt{\pi}}\sum_{n=1}^{\left[\frac{D}{2}\right]}\sum_{j=0}^{2n+1}\frac{\Gamma\left(j+n-\frac{1}{2}\right)}{\Gamma\left(j+n\right)}
  \Bigg[\textrm{FP}\,\zeta_{N}\left(n-\frac{1}{2}\right)\mathcal{C}_{j,n}(a,R)\nonumber\\
  &+&\textrm{Res}\,\zeta_{N}\left(n-\frac{1}{2}\right)\mathcal{F}_{j,n}(a,R)\Bigg]+\frac{1}{2}\left(\frac{1}{\varepsilon}+\ln\mu^{2}\right)\Bigg\{\frac{1}{4\pi}\zeta_{N}(-1)\int_{a}^{R}f^{-2}(t)\diff t\nonumber\\
  &+&\frac{1}{2\pi}\zeta_{N}(0)\int_{a}^{R}F_{0,1}(t)\diff t
  -\frac{1}{4}\textrm{Res}\,\zeta_{N}\left(-\frac{1}{2}\right)\left[f^{-1}(a)+f^{-1}(R)\right]\nonumber\\
  &+&\frac{1}{\sqrt{\pi}}\sum_{n=1}^{\left[\frac{D-1}{2}\right]}\textrm{Res}\,\zeta_{N}\left(n\right)\sum_{j=0}^{2n+2}\frac{\Gamma(j+n)}{\Gamma\left(j+n+\frac{1}{2}\right)}\mathcal{A}_{j,n}(a,R)\\
  &+&\frac{1}{\sqrt{\pi}}\sum_{n=1}^{\left[\frac{D}{2}\right]}\textrm{Res}\,\zeta_{N}\left(n-\frac{1}{2}\right)\sum_{j=0}^{2n+1}\frac{\Gamma\left(j+n-\frac{1}{2}\right)}{\Gamma\left(j+n\right)}\mathcal{C}_{j,n}(a,R)\Bigg\}
+O(\varepsilon)\nonumber
\end{eqnarray}
where the notation $[x]$ stands for the integer part of $x$ and, for convenience, we have defined the function
\begin{eqnarray}
  \mathcal{Q}(a,R)&=&\frac{1}{4\pi}\left[\zeta'_{N}(-1)+(\ln 4-1)\zeta_{N}(-1)\right]\int_{a}^{R}f^{-2}(t)\diff t\nonumber\\
  &+&\frac{1}{2\pi}\zeta_{N}(-1)\int_{a}^{R}f^{-2}(t)\ln f(t)\diff t\nonumber\\
  &-&\frac{1}{4}\textrm{FP}\,\zeta_{N}\left(-\frac{1}{2}\right)\left[f^{-1}(a)+f^{-1}(R)\right]+\frac{1}{2\pi}\zeta'_{N}(0)\int_{a}^{R}F_{0,1}(t)\diff t\nonumber\\
  &-&\frac{1}{2}\textrm{Res}\,\zeta_{N}\left(-\frac{1}{2}\right)\left[f^{-1}(a)\ln f(a)+f^{-1}(R)\ln f(R)\right]\nonumber
\end{eqnarray}
\begin{eqnarray}\label{45}
  &+&\frac{1}{2\pi}\zeta_{N}(0)\left[2\int_{a}^{R}F_{0,1}(t)\ln f(t)\diff t+(\ln 2-1)\int_{a}^{R}F_{0,1}(t)\diff t\right]\nonumber\\
  &-&\frac{1}{2\sqrt{\pi}}\zeta_{N}(0)\sum_{j=1}^{2}\frac{\Gamma(j)}{\Gamma\left(j+\frac{1}{2}\right)}\int_{a}^{R}F_{j,1}(t)f^{2j}(t)\diff t\;.
\end{eqnarray}
The Casimir energy in region $II$ can be obtained from (\ref{new}) once the replacement $a\to R$ and $R\to b$ is performed and by considering $Z_{II}(s,R,b)$ in place of $Z_{I}(s,a,R)$.

In order to compute the Casimir force on the piston we need, according to (\ref{41a}) and (\ref{10a}), the residue and finite part of the spectral zeta function $\zeta_{M}(s)$ at $s=-1/2$.
By adding the contributions coming from region $I$ and region $II$ we obtain, for the residue, the following expression
\begin{eqnarray}\label{43}
  \textrm{Res}\,\zeta_{M}\left(-\frac{1}{2}\right)&=&\frac{1}{4\pi}\zeta_{N}(-1)\int_{a}^{b}f^{-2}(t)\diff t+\frac{1}{2\pi}\zeta_{N}(0)\int_{a}^{b}F_{0,1}(t)\diff t
  \nonumber\\
  &-&\frac{1}{4}\textrm{Res}\,\zeta_{N}\left(-\frac{1}{2}\right)\left[f^{-1}(a)+f^{-1}(b)+2f^{-1}(R)\right]\nonumber\\
  &+&\frac{1}{\sqrt{\pi}}\sum_{n=1}^{\left[\frac{D-1}{2}\right]}\textrm{Res}\,\zeta_{N}\left(n\right)\sum_{j=0}^{2n+2}\frac{\Gamma(j+n)}{\Gamma\left(j+n+\frac{1}{2}\right)}\mathcal{A}_{j,n}(a,b)\nonumber\\
  &+&\frac{1}{\sqrt{\pi}}\sum_{n=1}^{\left[\frac{D}{2}\right]}\textrm{Res}\,\zeta_{N}\left(n-\frac{1}{2}\right)\sum_{j=0}^{2n+1}\frac{\Gamma\left(j+n-\frac{1}{2}\right)}{\Gamma\left(j+n\right)}\nonumber\\
  &\times&\left[\mathcal{C}_{j,n}(a,b)-\mathcal{T}_{j,n}(R)\right]\;.
\end{eqnarray}
Furthermore, for the finite part we have
\begin{eqnarray}\label{44}
  \textrm{FP}\,\zeta_{M}\left(-\frac{1}{2}\right)&=&Z_{I}\left(-\frac{1}{2},a,R\right)+Z_{II}\left(-\frac{1}{2},R,b\right)+\mathcal{Q}(a,b)+\mathcal{Y}(R)\nonumber\\
  &+&\frac{1}{\sqrt{\pi}}\sum_{n=1}^{\left[\frac{D-1}{2}\right]}\sum_{j=0}^{2n+2}\frac{\Gamma(j+n)}{\Gamma\left(j+n+\frac{1}{2}\right)}
  \big[\textrm{FP}\,\zeta_{N}\left(n\right)\mathcal{A}_{j,n}(a,b)\nonumber\\
  &+&\textrm{Res}\,\zeta_{N}\left(n\right)\mathcal{B}_{j,n}(a,b)\big]\nonumber\\
  &+&\frac{1}{\sqrt{\pi}}\sum_{n=1}^{\left[\frac{D}{2}\right]}\sum_{j=0}^{2n+1}\frac{\Gamma\left(j+n-\frac{1}{2}\right)}{\Gamma\left(j+n\right)}
  \Bigg[\textrm{FP}\,\zeta_{N}\left(n-\frac{1}{2}\right)\left(\mathcal{C}_{j,n}(a,b)-\mathcal{T}_{j,n}(R)\right)\nonumber\\
  &+&\textrm{Res}\,\zeta_{N}\left(n-\frac{1}{2}\right)\left(\mathcal{F}_{j,n}(a,b)-\mathcal{X}_{j,n}(R)\right)\Bigg],
\end{eqnarray}
where we have introduced the functions
\begin{equation}\label{46}
  \mathcal{Y}(R)=-\frac{1}{2}\textrm{FP}\,\zeta_{N}\left(-\frac{1}{2}\right)f^{-1}(R)-\textrm{Res}\,\zeta_{N}\left(-\frac{1}{2}\right)f^{-1}(R)\ln f(R)\;,
\end{equation}
and
\begin{equation}\label{47}
  \mathcal{X}_{j,n}(R)=\Omega_{j,n}(R)f^{2j+2n-1}(R)\left[\ln f(R)+\sum_{k=0}^{n+j-1}\frac{1}{2k-1}\right]\;.
\end{equation}

Thanks to the formulas (\ref{43}) and (\ref{44}) the Casimir force on the piston, which is the quantity of main interest, can be evaluated according to (\ref{10a}). By differentiating the above results with respect to the position $R$ of the piston and by denoting with a prime such derivative we obtain
\begin{eqnarray}\label{48}
  F_{\textrm{Cas}}(R)&=&-\frac{1}{2}Z'_{I}\left(-\frac{1}{2},a,R\right)-\frac{1}{2}Z'_{II}\left(-\frac{1}{2},R,b\right)-\frac{1}{2}\mathcal{Y}'(R)\nonumber\\
   &+&\frac{1}{2\sqrt{\pi}}\sum_{n=1}^{\left[\frac{D}{2}\right]}\sum_{j=0}^{2n+1}\frac{\Gamma\left(j+n-\frac{1}{2}\right)}{\Gamma\left(j+n\right)}
  \Bigg[\textrm{FP}\,\zeta_{N}\left(n-\frac{1}{2}\right)\mathcal{T}'_{j,n}(R)\nonumber\\
  &+&\textrm{Res}\,\zeta_{N}\left(n-\frac{1}{2}\right)\mathcal{X}'_{j,n}(R)\Bigg]-\frac{1}{4}\left(\frac{1}{\varepsilon}+\ln\mu^{2}\right)
  \Bigg\{\textrm{Res}\,\zeta_{N}\left(-\frac{1}{2}\right)\frac{f'(R)}{f^{2}(R)}\nonumber\\
  &-&\frac{2}{\sqrt{\pi}}\sum_{n=1}^{\left[\frac{D}{2}\right]}\textrm{Res}\,\zeta_{N}\left(n-\frac{1}{2}\right)\sum_{j=0}^{2n+1}\frac{\Gamma\left(j+n-\frac{1}{2}\right)}{\Gamma\left(j+n\right)}\mathcal{T}'_{j,n}(R)\Bigg\}+O(\varepsilon)\;,
\end{eqnarray}
with
\begin{equation}
  \mathcal{Y}'(R)=\frac{f'(R)}{f^{2}(R)}\left[\frac{1}{2}\textrm{FP}\,\zeta_{N}\left(-\frac{1}{2}\right)+\textrm{Res}\,\zeta_{N}\left(-\frac{1}{2}\right)(1-\ln f(R))\right]\;,
\end{equation}
\begin{equation}
  \mathcal{T}'_{j,n}(R)=\frac{1}{2}f^{2j+2n-2}(R)\left[\Omega'_{j,n}(R)f(R)+(2n+2j-1)\Omega_{j,n}(R)f'(R)\right]\;,
\end{equation}
and
\begin{equation}\label{49}
  \mathcal{X}'_{j,n}(R)=2\mathcal{T}'_{j,n}(R)\left[\ln f(R)+\sum_{k=0}^{n+j-1}\frac{1}{2k-1}\right]+2\mathcal{T}_{j,n}(R)\frac{f'(R)}{f(R)}\;.
\end{equation}
We would like to make a few remarks regarding the result (\ref{48}) for the Casimir force on the piston. The term proportional to $(1/\varepsilon+\ln\mu^{2})$ represents the ambiguity which is generally present in the Casimir force. The ambiguity is proportional to the residue of $\zeta_{N}(s)$ at the points $s=n-1/2$ with $0\leq n\leq [D/2]$. The residues
are related to the heat kernel coefficients $a_{D/2-n}$ associated with the manifold $N$ which shows that the ambiguity is purely of geometric nature. It is clear, from the previous remark, that
the force on the piston becomes a well defined quantity when the manifold $N$ is even-dimensional and without boundary, a result that also holds for Casimir pistons modeled by a generalized cone \cite{fucci11b,fucci11c}. We would like to point out that the term in expression (\ref{48}) responsible for the ambiguity of the force is proportional to the first and higher derivatives of the warping functions $f(R)$. This can be easily seen for the term proportional to $\textrm{Res}\,\zeta_{N}\left(-1/2\right)$. Moreover, for the remaining terms we notice that $\mathcal{T}'_{j,n}(R)$ contains first and higher derivatives of $f(R)$ through $\Omega_{j,n}(R)$ and $\Omega'_{j,n}(R)$ (see, for instance, Eqs. (\ref{25}) and (\ref{22})).
The last remark shows that when the warping function is constant, namely the warped manifold $M$ becomes a generalized cylinder, the force is always
unambiguous regardless of the geometry of the manifold $N$, as has been observed earlier \cite{mara}.

\section{Conclusions}

In this work we have studied the Casimir energy and force for massless scalar fields endowed with Dirichlet boundary conditions in the
framework of a generalized piston constructed from a warped product manifold. In order to compute the Casimir energy and force in this setting we have
performed the analytic continuation to a neighborhood of $s=-1/2$ of the spectral zeta function associated with the Laplace operator defined on warped product manifolds.
The method employed for the analytic continuation is based on the contour integral representation of the zeta function and the WKB expansion of the eigenfunctions of the Laplacian on the manifold $M$. This procedure allows one to study the spectral zeta function associated with manifolds possessing quite general geometries. A more complete and detailed account of the method has been presented in \cite{fucci11k}.

The results that we have obtained for the Casimir energy and force are valid in any dimension $D$ and for any smooth, compact piston $N$. In addition, both the expressions for the energy and force depend explicitely on the spectral zeta function $\zeta_{N}(s)$ which shows how these quantities are related to the geometry and the topology of the piston.
The interest in studying generalized piston configurations lies in the fact that, for arbitrary warping functions, the two chambers forming the piston do not have the same geometry.
This implies that the behavior of the Casimir force for generalized piston configurations is different from the usual Casimir pistons considered in the literature where the two chambers have the same geometry. We would like to mention that the formulas for the Casimir force on the piston, although very general, are somewhat implicit. However, once a specific warping function $f(r)$ and manifold $N$ are chosen more explicit results can be obtained.

In fact, this would be an interesting continuation of the presented research. Namely, study in more detail particular examples of generalized pistons which are also of relevance in physical situations. Furthermore other boundary conditions can be envisaged by essentially the same methods.\\[.3cm]

\noindent
{\bf Acknowledgments}\\
KK would like to thank the organizers for introducing a special session in honor of Stuart Dowker's 75th birthday into the conference schedule. In that session KK delivered a talk describing the life and work of Stuart. The details presented will be incorporated in the Introduction of a Journal of Physics A special issue in Stuart's honor published next year under the title ``Applications of zeta functions and other spectral functions in mathematics and physics". KK is supported by the National Science Foundation Grant PHY-0757791.

\end{document}